\documentclass[twocolumn,showpacs,preprintnumbers,amsmath,amssymb,prb,aps]{revtex4-2}

\usepackage{subfigure}
\usepackage{comment}
\usepackage{esvect}
\usepackage[usenames, dvipsnames]{color}
\usepackage{epsfig}
\usepackage{graphicx}
\usepackage{float} 
\usepackage{multirow}
\usepackage{gensymb}
\usepackage[colorlinks=true,linkcolor=red,citecolor=blue,urlcolor=blue]{hyperref}
\usepackage{multirow}
\usepackage{dcolumn}   
\usepackage{bm}        
\usepackage{ulem}
\begin{document}
	
\title{Giant spectral renormalization and complex hybridization physics in a Kondo lattice system, CeCuSb$_2$}
	
\author{Sawani Datta,$^1$ Ram Prakash Pandeya,$^1$ Arka Bikash Dey,$^2$ A. Gloskovskii,$^2$ C. Schlueter,$^2$ Thiago Peixoto,$^2$ Ankita Singh,$^1$ A. Thamizhavel,$^1$ and Kalobaran Maiti$^1$}
\altaffiliation{Corresponding author: kbmaiti@tifr.res.in}
\affiliation{$^1$Department of Condensed Matter Physics and Materials Science, Tata Institute of Fundamental Research, Homi Bhabha Road, Colaba, Mumbai-400005, India.\\
$^2$Deutsches Elektronen-Synchrotron DESY, 22607 Hamburg, Germany.}
	
	
\begin{abstract}
We investigate the electronic structure of a Kondo lattice system, CeCuSb$_2$ exhibiting significant mass enhancement and Kondo-type behavior. We observe multiple features in the hard $x$-ray photoemission spectra of Ce core levels due to strong final-state effects. The depth-resolved data exhibit a significant change in relative intensity of the features with the surface sensitivity of the probe. The extracted surface and bulk spectral functions are different and exhibit a Kondo-like feature at higher binding energies in addition to the well and poorly screened features. The core-level spectra of Sb exhibit huge and complex changes as a function of the surface sensitivity of the technique. The analysis of the experimental data suggests that the two non-equivalent Sb sites possess different electronic structures and in each category, the Sb layers close to the surface are different from the bulk ones. An increase in temperature influences the Ce-Sb hybridization significantly. The plasmon-excitation-induced loss features are also observed in all core level spectra. All these results reveal the importance of Ce-Sb hybridizations and indicate that the complex renormalization of Ce-Sb hybridization may be the reason for the exotic electronic properties of this system.
\end{abstract}
	
\maketitle
	
\section{Introduction}
In rare-earth intermetallics, the hybridization between the localised 4$f$ states and the itinerant states gives rise to several exotic phenomena like valence fluctuation,\cite{Shin_val.fluct,C1,C2} heavy fermionic behaviour, Kondo physics,\cite{patil1,maiti,patil2} unconventional superconductivity,\cite{wat_uncon.supercon} etc. The driving mechanism for such complex properties is often associated with competing interactions of the strong electron correlation of the $f$ states, crystal electric field,\cite{patil_CEF} spin-orbit coupling, etc. These interaction parameters are strongly dependent on the local environment around the rare-earth site as well as the kinetic energy of the conduction states which is derived by the hopping of charge carriers. To realize varied properties, these parameters are exploited efficiently via the change in lattice symmetries, chemical compositions, pressure, etc. The presence of one 4$f$ electron and its proximity to the Fermi level in Ce relative to other rare-earths makes Ce-based materials highly interesting.\cite{patil3} Strong hybridization of 4$f$ states with the conduction states leads to several unique properties of Ce-based compounds. This nature of Ce amused both experimentalists and theorists for many decades.

\begin{figure}
\centering
\includegraphics[width=0.5\textwidth]{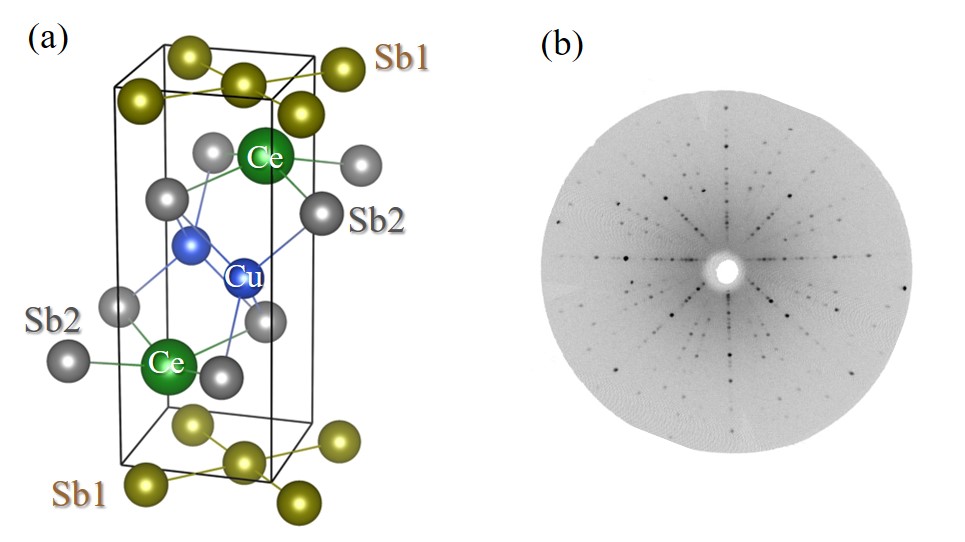}
\caption{(a) Crystal structure of CeCuSb$_2$. (b) Laue pattern exhibiting high crystallinity of the sample.
}
\label{structure}
\end{figure}
	
Among the varied class of materials, the electronic structure of ternary Ce-antimonides (CeTX$_2$; T = Cu, Au, Ag ..., X = Sb, As, Te ...) are relatively less studied. The uniqueness of these materials is that they form in a layered tetragonal ZrCuSi$_2$-type structure.\cite{layer_brylak,USb2_Kac} For example, in CeCuSb$_2$, the structure contains stacking of the Sb1-CeSb2-Cu-Sb2Ce-Sb1 layers as shown in Fig. \ref{structure}(a). Here, Sb1 and Sb2 are the two non-equivalent Sb atoms with different Wycoff positions. This crystal lattice belongs to the space group $P4/nmm$(129) with lattice constants, $a$ = 4.337 \AA\ and $c$ = 10.233 \AA.\cite{thamizh} In this structure, the Ce-atoms are more strongly hybridized with the conduction layers consisting of Sb2CuSb2 layers. Sb1 atoms form a square net structure and relatively weakly coupled to Ce.
These materials show various interesting phenomena like anomalous bulk properties, mixed-valency, heavy fermionic nature, Kondo effect, etc. For example, CeCuSb$_2$ and CeAuSb$_2$ are antiferromagnets with Ne\'{e}l temperatures of 6.9 K and 5.0 K, respectively but CeNiSb$_2$ is found to order ferromagnetically below a Curie temperature of 6.0 K.\cite{thamizh} On the other hand, CeCuAs$_2$ do not show magnetic order down to the lowest temperature studied.\cite{Sengupta1,Sengupta2}
	
Here, we studied the electronic structure of CeCuSb$_2$ using hard x-ray photoemission spectroscopy (HAXPES). The resistivity of CeCuSb$_2$ increases logarithmically with cooling from the room temperature, it becomes maximum around 23 K for the electric current, $J\parallel[100]$. For $J\parallel[001]$, the resistivity peak appears near 9 K manifesting strong anisotropy of the system.\cite{thamizh} Cooling below the Ne\'{e}l temperature decreases the resistivity as often observed in Kondo systems. From the magnetic measurements in the paramagnetic phase, the effective magnetic moment, $\mu_{eff}$ is found to be 2.32 $\mu_B$/Ce for $H\parallel$ [100], and 2.33 $\mu_B$/Ce for $H\parallel$ [001]. Such reduced $\mu_{eff}$ from the free Ce$^{3+}$ ion value of 2.54 $\mu_B$/Ce is interpreted as a signature of the Kondo effect as evidenced in the transport data.\cite{thamizh} The HAXPES data in this study show giant spectral renormalization due to the change in surface sensitivity and temperature.
	
\section{Experiment}
High quality single crystals of CeCuSb$_2$ were grown using Sb as self flux by high temperature solution growth method.\cite{thamizh} Highly pure constituent elements Ce, Cu and Sb were taken in the molar ratio of 1:2:21 in a high quality recrystallized alumina crucible; here, excess Cu was used to avoid the CeSb$_2$ phase formation.  The crucible containing the elements were sealed in an evacuated quartz ampoule with a partial pressure of Ar gas. The crucible was then placed in a box type resistive heating furnace. The heating was done slowly at the rate of 50~$^{\circ}$C/h up to 1050~$^{\circ}$C and held at this temperature for about 24 hours for proper homogenization. Then the furnace was cooled down at the rate of 2~$^{\circ}$C/hr down to 700~$^{\circ}$C, at which point the excess Sb flux was centrifuged.  Well defined shiny single crystals with platelet like morphology were obtained. The flat plane of the crystal corresponds to the (001) plane.

The chemical characterisation of the sample was done by the Energy dispersive analysis of $x$-rays and crystallinity of the sample was ensured by the Laue diffraction measurement. In Fig. \ref{structure}(b), a typical Laue pattern of CeCuSb$_2$ is shown. Sharp spots demonstrate the good crystallinity of the sample. The HAXPES measurements were carried out at the P22 beamline of PETRA III, DESY, Hamburg, Germany using a high-resolution Phoibos electron analyzer with an energy resolution close to 150 meV. The sample was cleaved in an ultra-high vacuum condition with the chamber pressure $\sim$10$^{-10}$ Torr and immediately transferred to the analysis chamber for the measurements where the base pressure was maintained below 10$^{-10}$ Torr. To vary the surface sensitivity of the technique, the measurements were done with three largely different photon energies, 8 keV, 6 keV, and 2.5 keV. The sample temperature was varied in the range of 45 K - 120 K using an open cycle helium cryostat.
	
\section{Results and discussion }
\begin{figure}
\centering
\includegraphics[width=0.5\textwidth]{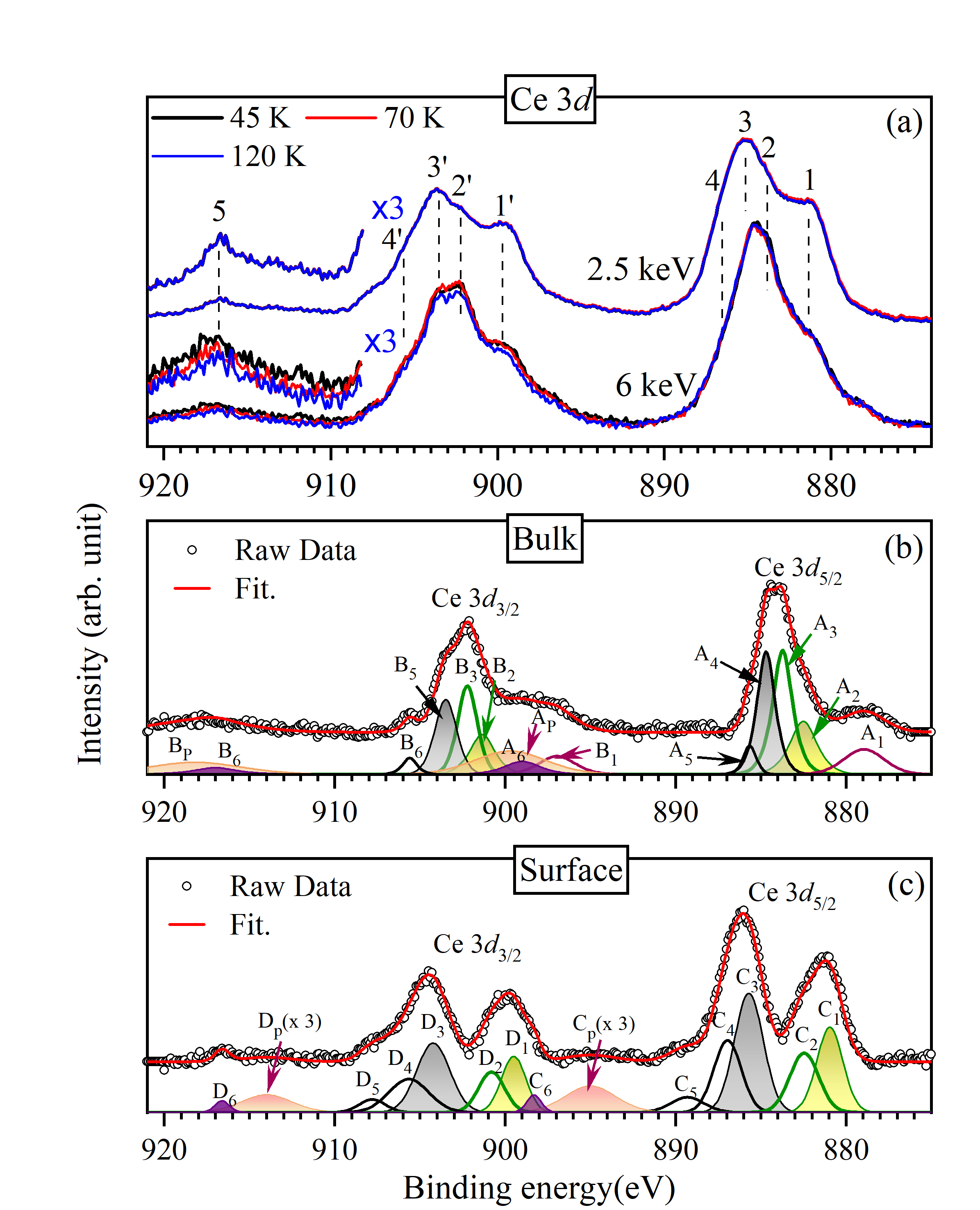}
\caption{(a) Temperature evolution of the Ce 3$d$ spectra collected at 45 K (black), 70 K (red) and 120 K (blue) using 2.5 keV and 6 keV photon energies. 1, 2... and 1$^\prime$, 2$^\prime$,... represent spin-orbit split 3$d_{5/2}$ and 3$d_{3/2}$ spectra respectively. Intensities beyond 910 eV in the 6 keV data are rescaled by 3 times to show the temperature dependence of the feature, 5. The raw data (black open circle) at 45 K and the fit curve (red line) of the (b) bulk and (c) surface Ce 3$d$ spectra are superimposed. The component peaks are shown in the lower panel. The spin-orbit split 3$d_{5/2}$ and 3$d_{3/2}$ spectra are denoted by respectively A's, B's in (b) and  C's, D's in (c). The intensity of the features A$_p$, B$_p$ in (b) and C$_p$, D$_p$ in (c) are enhanced by a factor of 2 and 3 respectively for clarity.
}
\label{Ce3d_temp}
\end{figure}

We study the Ce 3$d$ core-level spectra in Fig.~\ref{Ce3d_temp}. The experimental data were collected using 6 keV and 2.5 keV photon energies. In Fig.~\ref{Ce3d_temp}(a), the data at different temperatures are superimposed over each other. Each spectrum exhibits a large number of distinct features. The spin-orbit split 3$d_{5/2}$ peaks are denoted by 1, 2, etc., and the corresponding 3$d_{3/2}$ peaks are denoted by primed numbers, 1$^\prime$, 2$^\prime$, etc. In addition, there is an intense feature at higher binding energy which is denoted by 5. There are interesting changes in spectral features due to the change in photon energy, e.g. the intensity of the peak 1 and 3 (similarly 1$^\prime$ and 3$^\prime$) is larger in the 2.5 keV data. On the other hand, peak 2 and peak 4 are relatively more intense in the 6 keV spectra. The escape depth \cite{escape depth} of the Ce 3$d$ photoelectrons is about 36 \AA\ at 6 keV and hence represents essentially the bulk electronic structure. At 2.5 keV, the escape depth reduces to about 20 \AA\ and shows relatively more surface contributions than the 6 keV data. This suggests that peaks 1 and 3 are the surface features, and 2 and 4 are associated with the bulk electronic structure. Spectra collected at different temperatures are almost identical at both the photon energies except for a small increment of the feature 5 at lower temperatures in the 6 keV data.
	
In order to extract the surface and bulk spectral functions, we have taken the 45 K data collected using 6 keV and 2.5 keV. The photoemission spectral intensity $I(\epsilon)$ can be expressed as, $I(\epsilon)=\int_{0}^{d}I^s(\epsilon)e^{-x/\lambda}dx + \int_{d}^{\infty}I^b(\epsilon)e^{-x/\lambda}dx$. Here, $I^b(\epsilon)$ and $I^s(\epsilon)$ are the bulk and surface spectral functions, respectively. Effective thickness of the surface layer is $d$ and $\lambda$ is the escape depth of the photo-electron. Integration of the above equation provides, $I(\epsilon) = I^s(\epsilon)(1-e^{-d/\lambda}) + I^b(\epsilon)e^{-d/\lambda}$. Using this equation, $d/\lambda$ as a parameter and considering that the spectral intensity is non-negative, one can calculate the surface and bulk spectra from the 2.5 keV and 6 keV data.

The extracted bulk and surface spectra are shown by symbols in Fig. \ref{Ce3d_temp}(b) and (c), respectively. There are several distinct features in each case. Understanding the origin of these features requires a detailed calculation of the Ce 3$d$ core level spectra using the periodic Anderson model, which is beyond the scope of our study and may not be necessary for the conclusions of this manuscript. In order to get a qualitative idea about different final states in this case, we have simulated the surface and the bulk spectral functions by a set of Voigt functions. The peaks are considered for the distinct features observed in the experimental data. The least square error method is adopted to get the best fit. The intensity ratios of the spin-orbit split peaks are kept fixed according to the ratio of their degeneracy. The best fit results are shown by the red line superimposed over the experimental spectra in the figure.
	
Photoemission spectra represent the response functions of the excited states due to the incident photon. The eigenstates of the final state Hamiltonian which is different from the initial state Hamiltonian due to the presence of photo-hole will have finite transitions from the ground state. Therefore, multiple features are often observed in the photoemission spectra, this phenomenon is called the final state effects. The electronic configurations for various final states can be described as follow. (i) $f_2$ configuration corresponds to a {\it well-screened} final state where an electron from the neighboring sites hopped to the 4$f$ level at the photoemission site to screen the positive charge due to the 3$d$-hole.\cite{nickelates_kbm,RMP_Fujimori,cuprates_kbm} The presence of the well-screened feature indicates finite hybridization of the 4$f$ states with the valence states that enabled hopping.\cite{gun,imer} (ii) $f_1$-peak is known as a {\it poorly screened} feature where the core hole is not screened. Each of these configurations will consist of several multiplets due to the degeneracy of the 4$f$ levels and the availability of several ligand states. (iii) In the presence of 4$f$-hybridization with the valence states, the 4$f$ electrons can hop back and forth between the conduction band and 4$f$ level resulting in multiple-valency. In a strong coupling case, the local moment due to the 4$f$ electron can be compensated via the formation of a Kondo singlet. Thus, the 4$f$ electrons gain itineracy due to the entanglement with the conduction electrons. This state is often represented by a $f_0$-peak which appears at the higher binding energy side.\cite{gun,imer,f0-kondo,f0-kondo2,gun_f0}
	
We show the fitting results for the bulk spectra in Fig. \ref{Ce3d_temp}(b); the constituent peaks for the 3$d_{5/2}$ bulk photoemission are denoted by A's and 3$d_{3/2}$ peaks are denoted by B's. The spin-orbit splitting is found to be about 18.6 eV. Considering the binding energy of the features, the results from Gunnarsson-Sch\"{o}nhammer model\cite{gun} and subsequent study by Imer-Wuilloud,\cite{imer} we attribute the features (A$_2$, A$_3$) and (A$_4$, A$_5$) to the $f_2$ and $f_1$ features, respectively; multiple features represent the multiplet splitting of these strongly correlated states \cite{Pfau, Klein}. Such multiplets are observed in the Ce 3$d$ core level spectra of other systems.\cite{A6_ohno} The observation of a distinct feature, A$_1$ (B$_1$ in the 3$d_{3/2}$ case) is interesting and presumably related to the screening by highly delocalized electrons as observed in other systems.\cite{A1_Zhang,A1_bos,A1_ram,A1_manganite} These are observed essentially in the bulk spectra due to higher itineracy in the bulk as expected. In Fig. \ref{Ce3d_temp}(c), we show the experimental (black open circle) and simulated surface spectra (red lines). The 3$d_{5/2}$ peaks are denoted by C's and 3$d_{3/2}$ peaks by D's. Spin-orbit splitting is about 18.5 eV. Similar to the bulk spectra, C$_1$, C$_2$ (D$_1$ and D$_2$) are attributed to the multiplets of the well-screened, $f_2$ peaks and C$_3$, C$_4$, C$_5$ (D$_3$,C$_4$ and D$_5$) correspond to the multiplets of the poorly screened, $f_1$ peaks.
	
We observe two additional features in both the surface and bulk electronic structures. The features with the subscript, 'P' indicate plasmon excitations along with the core level photoemission; this attribution is discussed later in the text. The features A$_6$ and B$_6$ in the bulk spectra, and C$_6$ and D$_6$ in the surface spectra are very similar to the $f_0$ peak observed in earlier studies representing the signature of the Kondo peak.\cite{gun} The intensity of the $f_0$ peak appears to be very small in both the cases.
It is to be noted here that the peak positions in the surface Ce 3$d$ spectra ($f_1$ peak at 886 eV and $f_2$ peak at 881 eV) are very similar to the earlier reported data of trivalent Ce.\cite{Pfau} This is consistent with the observation of weaker $f_0$-peak in the surface spectra. Thus, it appears that the 4$f$-occupancy of the surface Ce is higher than the bulk 4$f$-occupancy.

\begin{figure}
\centering
\includegraphics[width=0.5\textwidth]{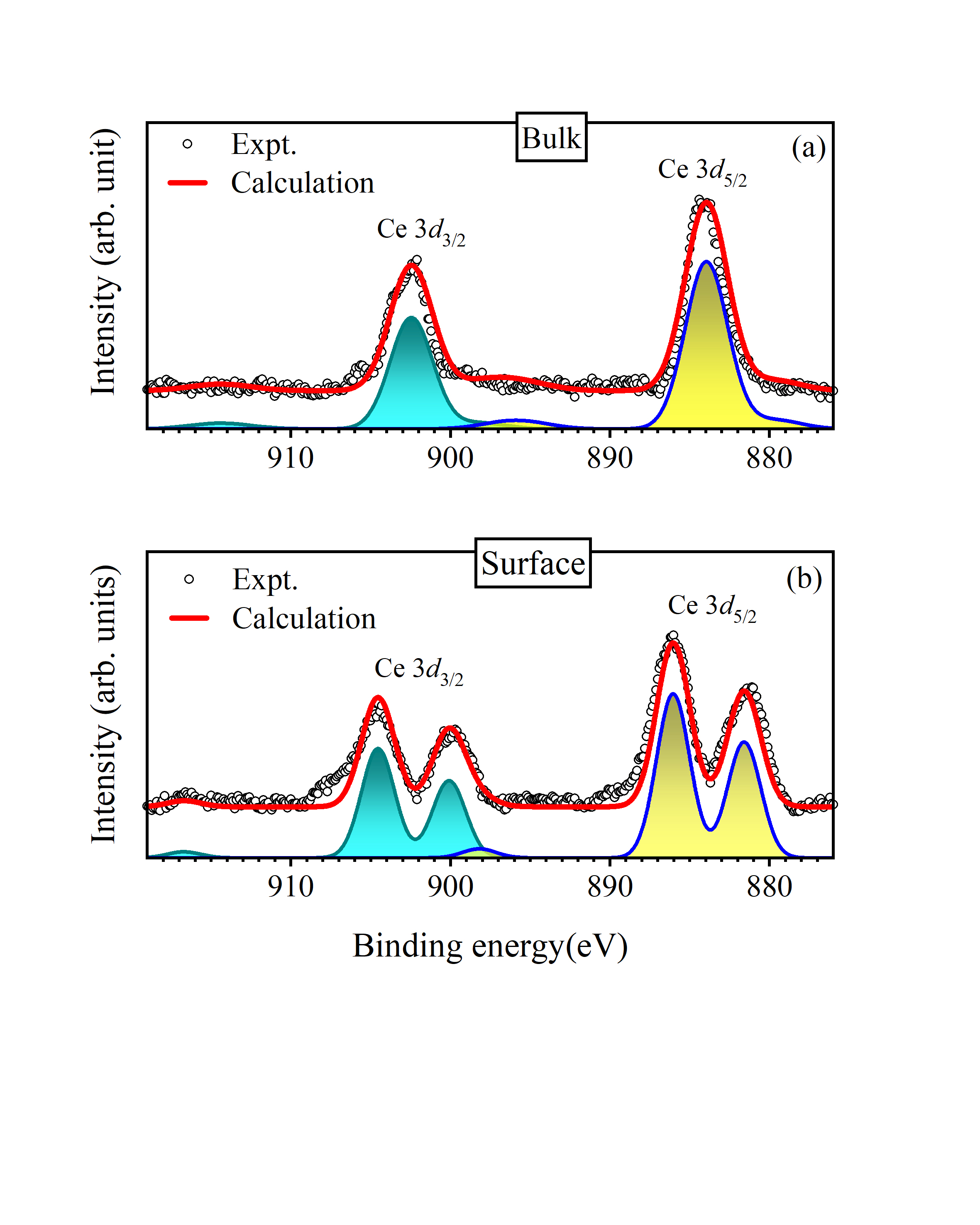}
\vspace{-18ex}
\caption{Experimental Ce 3$d$ (a) bulk and (b) surface spectral features to be simulated using Imer-Wuilloud configuration interaction model. Lines superimposed over the data are the simulated results. The area plots in the lower panel in each case show the spin-orbit split components.}
\label{ImerWuilloud}
\end{figure}

In order to verify this conclusion, we have simulated the surface and bulk spectral features following Imer-Wuilloud model.\cite{imer} The features, A$_1$ and B$_1$ are excluded from the simulation of the bulk spectrum as they are not considered in the model.
The simulated results are shown in Fig. \ref{ImerWuilloud}. While the detailed multiplet structures cannot be captured within this configuration interaction model, the simulated results shown by solid line superimposed over the experimental data exhibit a good description for the 4$f$ electron correlation strength, $U_{ff}$ = 9 eV and the charge transfer energy, $\Delta$ = 1 eV. The correlation strength between a 4$f$-electron and a core hole, $U_{fc}$ is found to be 10 eV. These parameters are same for both the surface and bulk spectra. The binding energy of the Ce 4$f$ electrons, $\epsilon_f$ is found to be -2.3 eV for the bulk and -2.5 eV for the surface spectra. The 4$f$-occupancy estimated for the surface electronic structure is 0.99 which suggests that the surface Ce atoms possess close to trivalent configuration. The occupancy of bulk Ce 4$f$ is about 0.93 indicating an enhancement of Ce-valency in the bulk consistent with the experimental results. It is to note here that the electron correlation strengths required to capture the experimental results using this model are somewhat larger than typical values found in Ce-based compounds. More detailed calculations considering multiplet interactions are required to capture the experimental features. The present results indicate that the surface Ce valency is indeed close to trivalent configuration \cite{Pfau} and there is an enhancement of Ce-valency in the bulk electronic structure.

\begin{figure}
\centering
\includegraphics[width=0.5\textwidth]{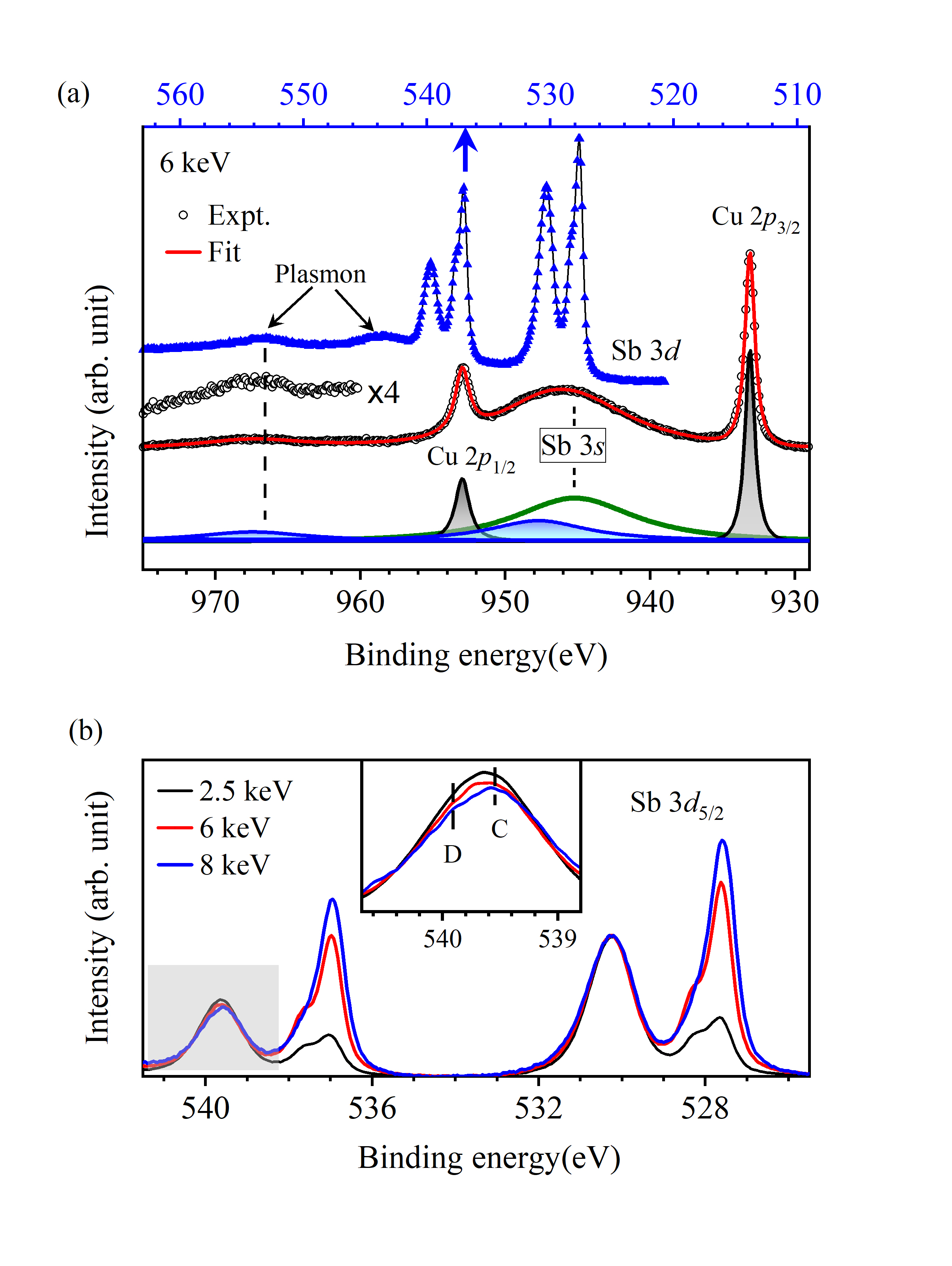}
\caption{(a) Sb 3$d$ ($x$-scale at the top) and Cu 2$p$ (bottom $x$-scale) core level spectra collected using 6 keV photon energy at 45 K. The line superimposed over the Cu 2$p$ spectrum is the fit data. The constituent peaks of the Cu 2$p$ spectral region are shown by lines in the lower panel. (b) Sb 3$d$ peaks at 45 K collected using varied photon energy. The inset shows the peak asymmetry near 539-540.5 eV binding energy region in an expanded energy scale.}
\label{Cu2pSb3d}
\end{figure}
	
In Fig.~\ref{Cu2pSb3d}(a), the Cu 2$p$ and Sb 3$d$ core level spectra are shown; the data were collected at 45 K using 6 keV photon energy. Sb 3$d$-spectra exhibit four intense features corresponding to 3$d_{5/2}$ and 3$d_{3/2}$ photoemission. The high binding energy regions of the Sb 3$d$ signal show two distinct features appearing about 14 eV away from the 3$d_{5/2}$ and 3$d_{3/2}$ peaks; the energy separation of the features is identical to the spin-orbit splitting of Sb 3$d$ levels. Cu 2$p$ spectrum also shows the signature of such features; the higher binding energy region is shown by rescaling the intensity for clarity. Similar features are also observed in the Ce 3$d$ spectra denoted by the subscript, P in Fig. \ref{Ce3d_temp}. The appearance of such features in all the core level excitations indicates their link to the energy loss due to the excitation of collective modes. The energy scale of these loss features suggests the presence of electron-plasmon coupling in this system.
	
Cu 2$p$ spectra show two sharp peaks and a broad hump in between them. To identify the constituting features, we have simulated the spectral function considering Voigt functions for the distinct peaks. The fitted envelope is shown by the red line superimposed over the experimental data. The intensity in between the Cu 2$p_{3/2}$ and 2$p_{1/2}$ peak appears due to the Sb 3$s$ excitations as marked in the figure. It is clear that Cu 2$p$ spin-orbit split contributions can be simulated by a single peak structure without satellite contributions. The binding energies of these peaks are about 953 eV and 933 eV; the spin-orbit splitting is about 20 eV. These results suggest the monovalency of Cu in CeCuSb$_2$ \cite{chainani}.
	
In Fig.~\ref{Cu2pSb3d}(b), we analyse the Sb 3$d$ core level peaks probed by different photon energies at 45 K. Each of the spin-orbit split Sb 3$d$ (3$d_{5/2}$ and Sb 3$d_{3/2}$) spectral regions show multiple peaks. For example, Sb 3$d_{3/2}$ spectral region show distinct peaks at 537, 537.7 and 539.5 eV. The intensity of the features changes significantly with the change in photon energy; all the spectra are normalized by the intensity at 530.3 eV. Such huge spectral changes with photon energy suggest significant surface-bulk differences as the change in photon energy essentially changes the surface sensitivity of the technique. The feature around 539.5 eV also shows some spectral evolution; this is shown in the inset of the figure. Clearly, there are two peaks in this region as marked by C and D. With the increase in photon energy, the intensity of D reduces with respect to the intensity of C leading to an overall change in spectral lineshape.
	
\begin{figure}
\centering
\includegraphics[width=0.5\textwidth]{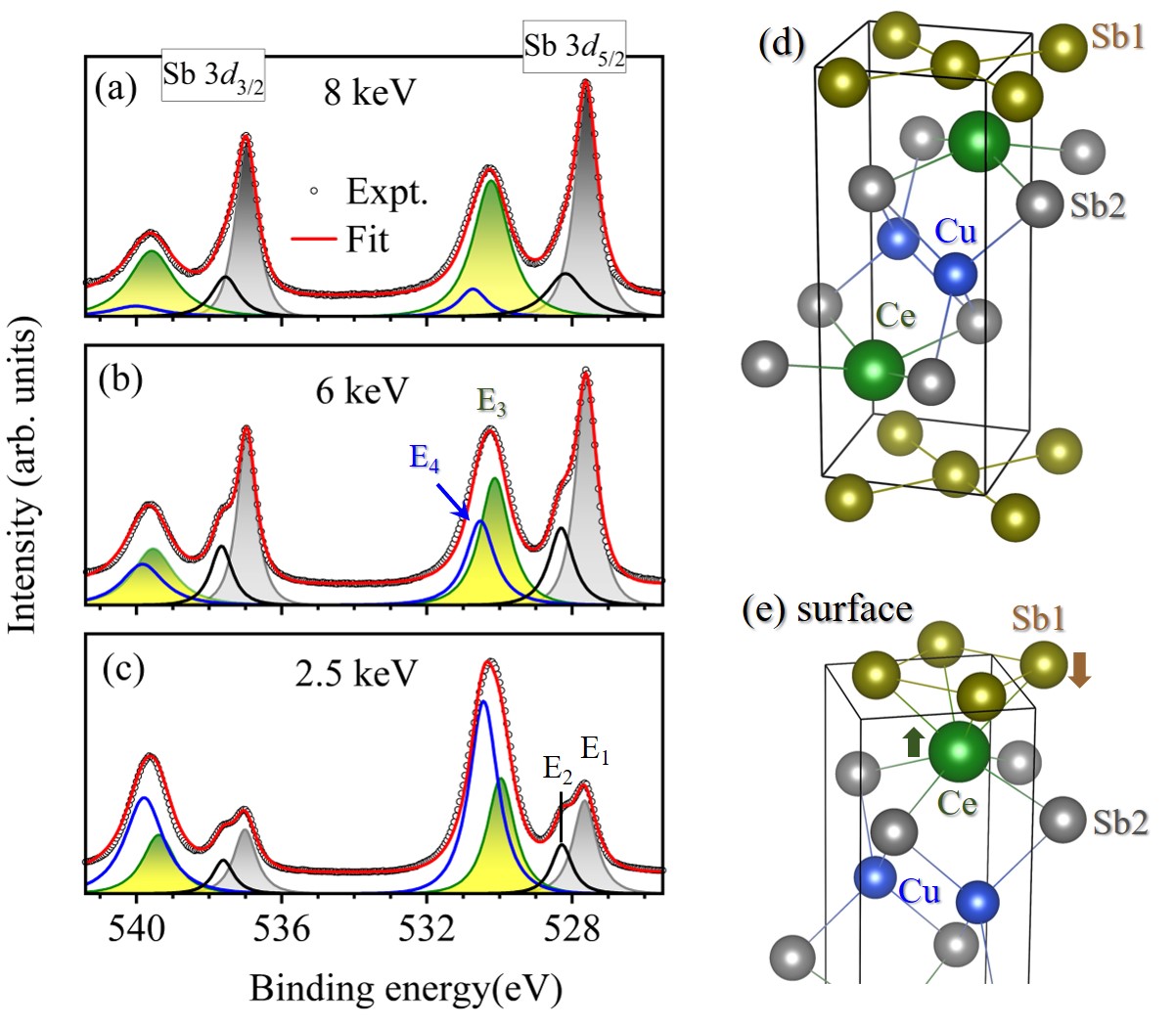}
\caption{Simulation of the Sb 3$d$ spectra collected at 45 K using (a) 8 keV, (b) 6 keV and (c) 2.5 keV photon energies. The open circles represent the experimental data and the lines superimposed over them are the fit data. The lines below the experimental data represent the constituent peaks. (d) Bulk unit cell of CeCuSb$_2$. (e) Schematic of the surface unit cell in the ground state; surface termination renormalizes Ce-Sb1 and Ce-Sb2 bondlengths.}
\label{Sb3dfit}
\end{figure}

To analyse the Sb 3$d$ spectra in more detail, we have fit the experimental spectra as shown in Fig.~\ref{Sb3dfit}. Fig.~\ref{Sb3dfit}(a), (b) and (c) show the Sb 3$d$ spectra recorded with 8 keV, 6 keV and 2.5 keV photon energies. The gray shaded peaks represent the lowest energy spin-orbit split features and appear at the binding energies of 527.6 eV and 537 eV; the spin-orbit splitting is about 9.4 eV. In each of the spin-orbit split spectral region, there are four peaks; for 3$d_{5/2}$ features, they are denoted by $E_1$, $E_2$, $E_3$ and $E_4$. $E_2$, $E_3$ and $E_4$ are about 0.7, 2.5, and 3 eV away from the lowest energy feature, $E_1$. With the decrease in photon energy, spectral features exhibit drastic changes; (i) the relative intensity in the region 526-529 eV with respect to the intensity between 529-532 eV reduces and (ii) the relative intensity ratios, $E_2/E_1$ and $E_4/E_3$ enhances gradually.
	
	
In order to get a qualitative understanding of the above scenario, we look at the crystal structure of CeCuSb$_2$ in Fig.~\ref{Sb3dfit}(d). The unit cell is shown by solid lines. The crystal structure is a layered structure and there are two nonequivalent positions of Sb as denoted by Sb1 and Sb2. Cu layer is sandwiched by two Sb2 layers, which indicates a strong hybridization between the Cu-Sb2. Thus, the core hole in Sb2 will be efficiently screened by the electron density in Cu leading to a feature at lower binding energies. Also, the Ce-Sb1 bond is longer (3.34586 \AA) than the Ce-Sb2 bond length (3.22972 \AA). Hence, Sb1 is more atomic-like having weaker hybridization with Ce compared to Ce-Sb2 hybridization. The reported density of states calculation of this sample supports the greater hybridization of Sb2 5$p$ state with Ce 5$d$ and Cu 3$d$ states.\cite{CeCuSb2} In addition, the Sb2 5$p$ occupancy in the band structure results is significantly larger than Sb1 5$p$ occupancy suggesting Sb2 valency to be more negative than Sb1 valency.
Therefore, the higher binding energy features $E_3$ and $E_4$ are attributed to the photoemission from Sb1 sites and the lower binding energy features $E_1$ and $E_2$ represent the Sb2 spectra. The large intensity of $E_3$ and $E_4$ in the 2.5 keV data and its reduction in intensity with the increase in bulk sensitivity of the technique suggests that the sample is cleaved at the Sb1 layer.
	
\begin{figure}
\centering
\includegraphics[width=0.5\textwidth]{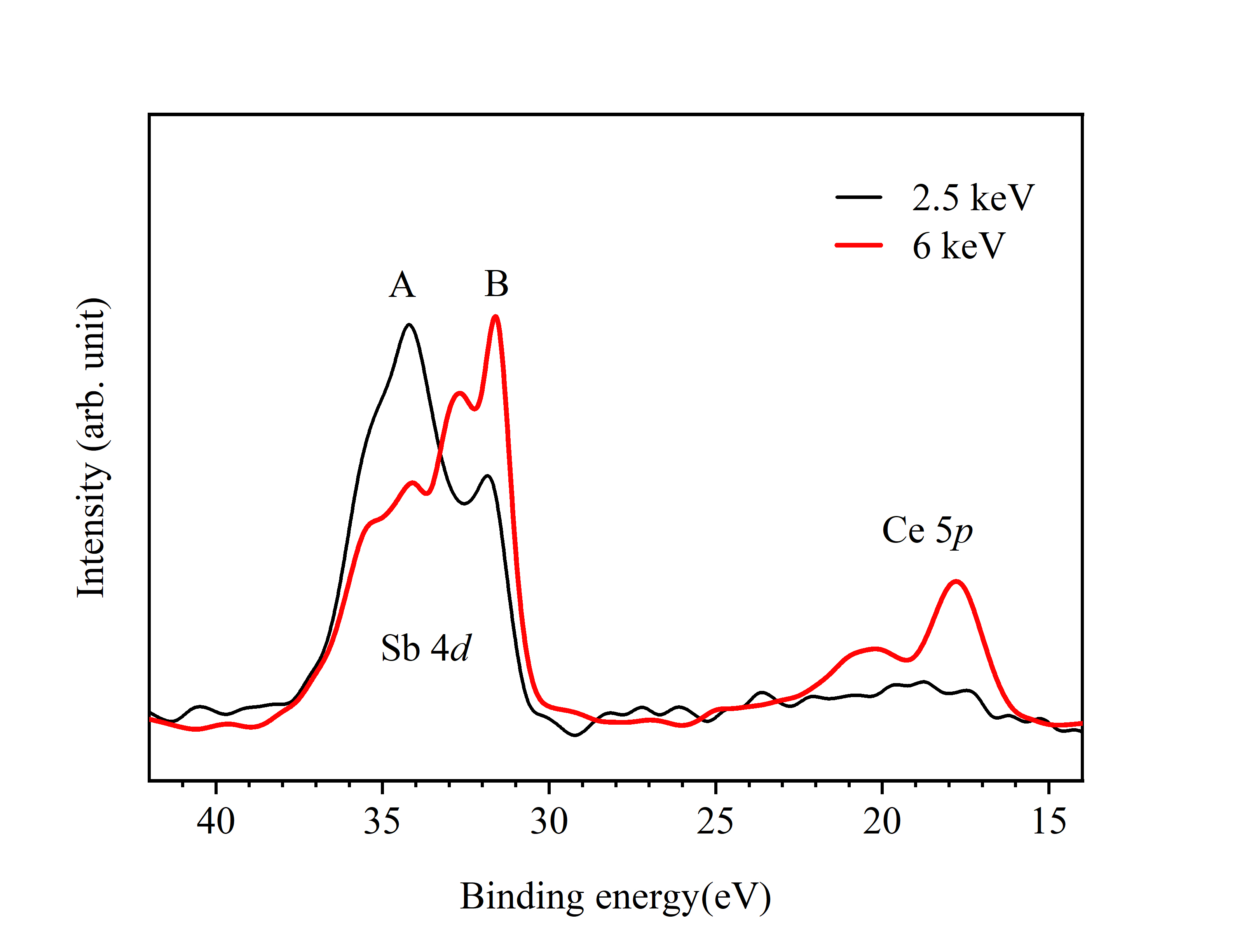}
\caption{Sb 4$d$ and Ce 5$p$ spectra at different photon energies.}
\label{core}
\end{figure}

In order to verify this identification, we show the Ce 5$p$ and Sb 4$d$ core level spectra collected at different photon energies in Fig. \ref{core}. It is evident from the figure that the intensity of Ce 5$p$ is the weakest in the most surface sensitive case of 2.5 keV data. The relative Ce 5$p$ intensity is enhanced significantly in the more bulk sensitive 6 keV data suggesting that the surface electronic structure is dominated by Sb contributions; the relative photoemission cross-section remains almost the same at these two energies. Within the Sb 4$d$ spectral region, we observe two sets of features marked as A and B in the figure. The intensity of A is highest in the 2.5 keV data which can be attributed to the surface Sb contributions which is Sb1 if the cleaving occurs at Sb-layer. The feature, A becomes less intense along with an enhancement of feature B intensity in the 6 keV data. These results establish the description of cleaving of the sample at the Sb1 layer discussed above as well as the identification of Sb1 and Sb2 features.

The scenario at the surface will be significantly different from the bulk as discussed in Fig. \ref{Sb3dfit}(e). The breaking of translational symmetry at the surface will lead to a distortion in the Ce-Sb2 layer which will change the relative Ce-Sb2 and Ce-Sb1 bond lengths as shown by arrows in the figure. Therefore, the bonding of the surface Sb1 layer with the Ce-layer underneath may become relatively stronger than the Sb1 layers in the bulk affecting Ce-Sb2 bonding near the surface layers. In any case, the surface Sb1 layer will be most atomic-like due to the absence of Ce-layers on top arising from translational symmetry breaking. From the spectral changes with the photon energy, it is clear that the intensity of $E_2$ becomes less significant with the increase in bulk sensitivity and hence this can be attributed to the Sb2 layer close to the surface while the feature $E_1$ represent the bulk Sb2 signal. A similar scenario is observed in the case of $E_3$ and $E_4$; $E_4$ is most intense in the 2.5 keV data and represents the spectra corresponding to the top Sb1 layer while the feature $E_3$ is the Sb1 layer within the bulk.
	
We note here that the Ce core level spectra show the signature of trivalent Ce-like behavior at the surface while the bulk Ce valency is larger. Thus, Sb2 close to the surface layer will be less hybridized with Ce compared to those in the bulk, and the effective Madelung potential at the surface Sb2 sites will be larger than that in the bulk. This explains qualitatively the two peak structures of the Sb2 3$d$ core-level spectra and the relative change in intensity of the two features with the change in surface sensitivity of the technique. We hope, these results will help to initiate detailed theoretical calculation in the future to understand the interesting electronic properties of such materials.

\begin{figure}
\centering
\includegraphics[width=0.45\textwidth]{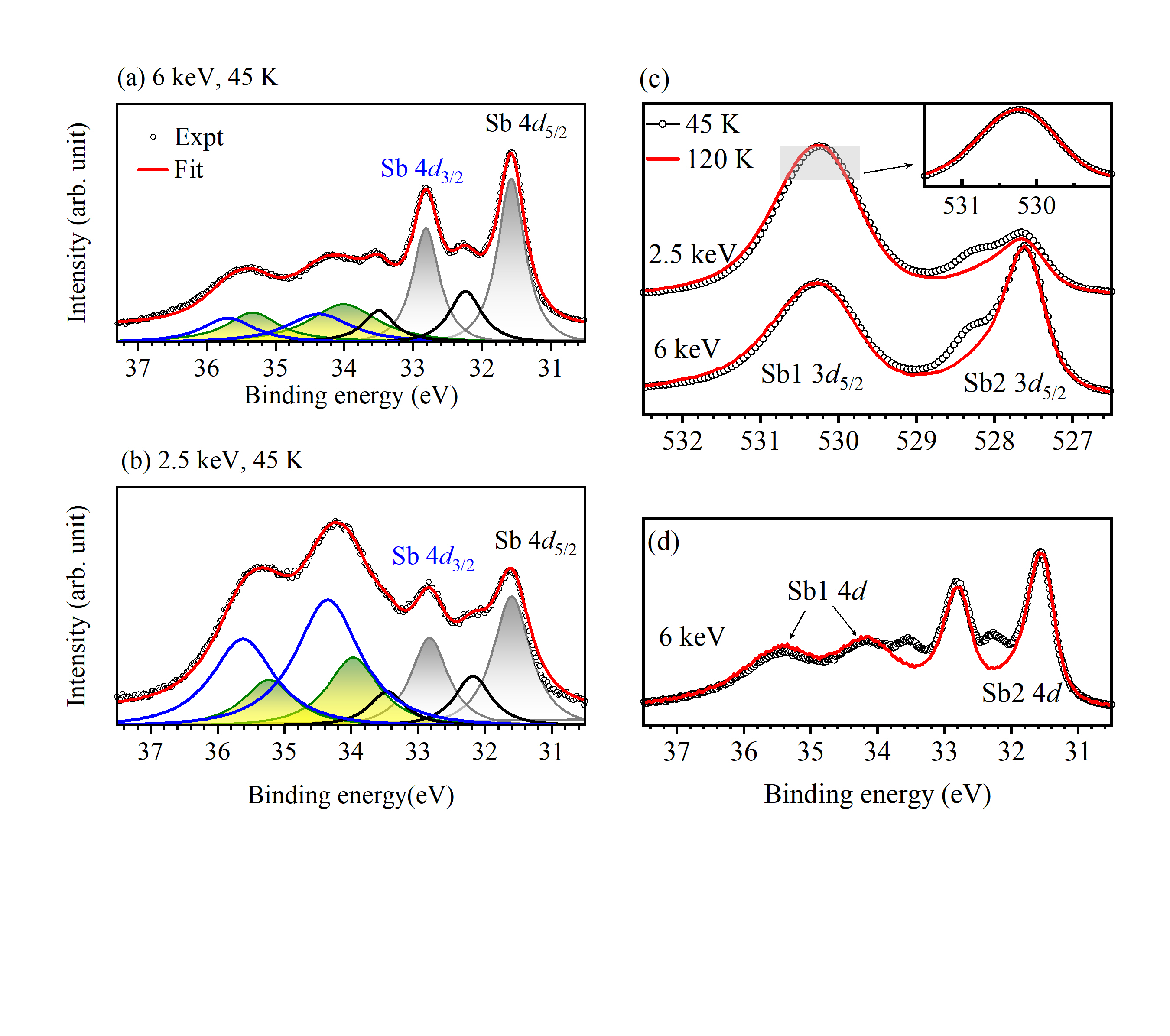}
\vspace{-12ex}
\caption{Sb 4$d$ spectra collected at 45 K using (a) 6 keV and (b) 2.5 keV photon energies. The experimental data are shown by symbols. The sumulation of these spectra are shown by lines. The line superimposed over the experimental data represent the total simulated spectral function and provide a good description. (c) Sb 3$d$ spectra collected at 45 K (symbols) and 120 K (lines) using 2.5 keV and 6 keV excitation energies. The inset shows 120 K data (line) superimposed on 45 K data (open circles) after shifting by 30 meV towards lower binding energy. (d) Temperature evolution of the 6 keV data of Sb 4$d$.}
\label{Sb4dfit}
\end{figure}
	
In order to investigate the consistency of the above description, we probed the Sb 4$d$ core level spectra. The experimental results collected at 45 K using 6 keV photon energy are shown in Fig. \ref{Sb4dfit}(a) and those collected with 2.5 keV photon energy are shown in  Fig. \ref{Sb4dfit}(b). There are clearly two sets of peaks each consisting of four features as found in Sb 3$d$ spectra. The intensity of these sets changes with the change in photon energy that identifies their surface and bulk character. Following the discussion above, the lower binding energy features are attributed to the photoemission signals from Sb2 while the ones are at higher binding energies are due to Sb1. The spin-orbit split spectra situated at about 31.6 eV (4d$_{5/2}$) and 32.8 eV (4d$_{3/2}$) binding energies are due to Sb2; the spin-orbit splitting is about 1.2 eV. In each spectral region, the lower binding energy features correspond to the bulk electronic structure and the surface features appear at higher binding energies.
	
We now turn to the temperature dependence of the spectral features. The Sb 3$d_{5/2}$ spectra at 2.5 keV photon energy is shown in Fig.~\ref{Sb4dfit}(c) for the sample temperatures of 45 K and 120 K. Sb1 peaks do not show significant evolution except a small shift of 30 meV towards higher binding energy at 120 K. This is shown in the inset by shifting the spectra by 30 meV and they overlap almost perfectly. The Sb2 features exhibit significant change; the distinct surface Sb2 feature is reduced significantly in the 120 K data. This is verified in the spectra collected at 6 keV [see Fig.~\ref{Sb4dfit}(c)] as well as Sb 4$d$ at 6 keV [Fig.~\ref{Sb4dfit}(d)]. All the spectra show identical scenarios indicating the disappearance of the distinct Sb2 surface feature at the higher temperature. This suggests that the structural change close to the surface discussed in Fig. \ref{Sb3dfit}(e) is relaxed due to thermal effects giving rise to almost identical properties of the Sb2 sites in the spectra. The thermal expansion of the Ce-Sb1 bond length would give rise to further weakening of the Ce-Sb1 hybridization which is reflected by a small shift of the Sb1 peak towards higher binding energy.
	
\section{Conclusion}
In conclusion, we have studied the electronic structure of a Kondo lattice system CeCuSb$_2$ using hard $x$-ray photoemission spectroscopy by varying the photon energy and the sample temperature. Depth-resolved studies reveal significantly different surface and bulk electronic structures. The results for Ce core level spectra indicate that the surface Ce has a lower valency than the valency of bulk Ce. We observe the evidence of Kondo-like feature in the Ce 3$d$ core-level spectra consistent with significant mass enhancement ($\gamma$ = 100 mJ/K$^2$) and Kondo-like resistivity upturn. Cu is found to be monovalent as observed in similar other systems. The core-level spectra of Sb exhibit giant changes with the change in photon energy and temperatures. There are two non-equivalent Sb sites in the crystal structure; Sb1 and Sb2. Analysis of the results suggests that the sample is cleaved at the Sb1 layer. The properties of the Sb layers close to the surface are very different from the bulk. The surface-bulk difference becomes less significant at higher temperatures. These experimental results suggest the presence of significant Ce-Sb hybridizations and their complex evolution with temperature which might be the reason for exotic electronic properties in this system.

\section{Acknowledgements}
Authors acknowledge the financial support under India-DESY program and Department of Atomic Energy (DAE), Govt. of India (Project Identification no. RTI4003, DAE OM no. 1303/2/2019/R\&D-II/DAE/2079 dated 11.02.2020). K.M. acknowledges support from BRNS, DAE, Govt. of India under the DAE-SRC-OI Award (grant no. 21/08/2015-BRNS/10977).


\begin{thebibliography}{11}
	
\bibitem{Shin_val.fluct}
S. Watanabe and K. Miyake, J. Phys.: Condens. Matter \textbf{23} 094217 (2011).
		
\bibitem{C1}
C. Laubschat, E. Weschke, M. Domke, C. T. Simmons, and G. Kaindl, Surface Science \textbf{269-270}, 605-609 (1992).
		
\bibitem{C2}
C. Laubschat, E. Weschke, C. Holtz, M. Domke, O. Strebel, and G. Kaindl,  Phys. Rev. Lett. \textbf{65}, 1639 (1990).
		
\bibitem{patil1}
S. Patil, S. K. Pandey, V. R. R. Medicherla, R. S. Singh, R. Bindu, E. V. Sampathkumaran, and K. Maiti, J. Phys.: Condens. Matter \textbf{22} 255602 (2010).
		
\bibitem{maiti}
K. Maiti, S. Patil, G. Adhikary, and G. Balakrishnan, J. Phys.: Conf. Ser. \textbf{273}, 012042(2011).
		
\bibitem{patil2}
S. Patil, G. Adhikary, G. Balakrishnan, and K. Maiti, J. Phys.: Condens. Matter \textbf{23}, 495601 (2011).
		
\bibitem{wat_uncon.supercon}
S. Watanabe and K. Miyake, J. Phys.: Condens. Matter \textbf{23}, 094217 (2011).
		
\bibitem{patil_CEF}
S. Patil, A. Generalov, M. Güttler, P. Kushwaha, A. Chikina, K. Kummer, T. C. R\"{o}del, A. F. Santander-Syro, N. Caroca-Canales, C. Geibel, S. Danzenb\"{a}cher, Yu. Kucherenko, C. Laubschat, J. W. Allen, and D. V. Vyalikh ,  Nat. Commun. \textbf{7} 11029 (2016).
		
\bibitem{patil3}
S. Patil, V. R. R. Medicherla, R. S. Singh, S. K. Pandey, E. V. Sampathkumaran, and K. Maiti, Phys. Rev. B \textbf{82}, 104428 (2010).
		
\bibitem{layer_brylak}
M. Brylak, M. H. M\"{o}ller, W. Jeitschko, J. Solid State Chem \textbf{115}, 305-308 (1995).
		
\bibitem{USb2_Kac}
D. Kaczorowski, R. Kruk, J. P. Sanchez, B. Malaman, and F. Wastin, Phys. Rev. B \textbf{58}, 9227 (1998).
		
\bibitem{thamizh}
A. Thamizhavel, T. Takeuchi, T. Okubo, M. Yamada, R. Asai, S. Kirita, A. Galatanu, E. Yamamoto, T. Ebihara, Y. Inada, R. Settai, and Y. $\bar{o}$nuki, Phys. Rev. B \textbf{68}, 054427 (2003).
		
\bibitem{Sengupta1}
K. Sengupta, E. V. Sampathkumaran, T. Nakano, M. Hedo, M. Abliz, N. Fujiwara, Y. Uwatoko, S. Rayaprol, K. Shigetoh, T. Takabatake, Th. Doert, and J. P. F. Jemetio, Phys. Rev. B \textbf{70}, 064406 (2004).
		
\bibitem{Sengupta2}
K. Sengupta S.Rayaprol, E.V.Sampathkumaran, Th. Doert, J.P.F.Jemetio, Physica B \textbf{348}, 465-474 (2004).
		
\bibitem{escape depth}
Yeh and Lindau, Atomic Data And Nuclear Data Tables \textbf{32}, l-l55 (1985).

\bibitem{nickelates_kbm}
K. Maiti, P. Mahadevan, and D. D. Sarma, Phys. Rev. B {\bf 59}, 12457 (1999).
		
\bibitem{RMP_Fujimori}
M. Imada, A. Fujimori, and Y.Tokura, Rev. Mod. Phys. \textbf{70}, 1039 (1998).
		
\bibitem{cuprates_kbm}
K. Maiti, D. D. Sarma, T. Mizokawa, and A. Fujimori, Phys. Rev. B {\bf 57}, 1572 (1998); K. Okada, A. Kotani, K. Maiti, and D. D. Sarma, J. Phys. Soc. Jpn. {\bf 65}, 1844 (1996); K. Maiti, D. D. Sarma, T. Mizokawa and A. Fujimori, Europhys. Lett. {\bf 37}, 359 (1997).

\bibitem{gun}
O. Gunnarsson and K. Sch\"{o}nhammer, Phys. Rev. B \textbf{28}, 4315 (1983).

\bibitem{imer}
J.-M. Imer and E. Wuilloud,  Z. Phys. B - Condensed Matter \textbf{66}, 153 (1987).

\bibitem{f0-kondo}
L. Braicovich, N. B. Brookes, C. Dallera, M. Salvietti, and G. L. Olcese, Phys. Rev. B, \textbf{56}, 15047(1997).
		
\bibitem{f0-kondo2}
J.C.Fuggle, F. U. Hillebrecht, Z. Zo\l{}nierek, R. L\"{a}sser, Ch. Freiburg, O. Gunnarsson, and K. Sch\"{o}nhammer, Phys. Rev. B, \textbf{27}, 7330 (1983).
		
\bibitem{gun_f0}
O. Gunnarsson and K. Sch\"{o}nhammer, Phys. Rev. B \textbf{31}, 4815 (1985).
		
\bibitem{Pfau}
A. Pfau and K. D. Schierbaum, Surface Science \textbf{321},71-80 (1994)
		
\bibitem{Klein}
M. Klein,  J. Kroha, H. v. L\"{o}hneysen, O. Stockert, and F. Reinert, Phys. Rev. B \textbf{79}, 075111 (2009)
		
		
\bibitem{A6_ohno}
Y. Ohno, Phys. Rev. B {\bf 48}, 5515 (1993).
		
\bibitem{A1_Zhang}
F. C. Zhang and T. M. Rice, Phys. Rev. B {\bf 37}, 3759(R) (1988).
		
\bibitem{A1_bos}
T. B\"{o}ske, K. Maiti, O. Knauff, K. Ruck, M. S. Golden, G. Krabbes, J. Fink, T. Osafune, N. Motoyama, H. Eisaki, and S. Uchida, Phys. Rev. B \textbf{57}, 138 (1998).
		
\bibitem{A1_ram}
R. P. Pandeya A. P. Sakhya, S. Datta, T. Saha, G. D. Ninno, R. Mondal, C. Schlueter, A. Gloskovskii, P. Moras, M. Jugovac, C. Carbone, A. Thamizhavel, and K. Maiti, Phys. Rev. B {\bf 104}, 094508 (2021).
		
\bibitem{A1_manganite}
K. Horiba, M. Taguchi, A. Chainani, Y. Takata, E. Ikenaga, D. Miwa, Y. Nishino, K. Tamasaku, M. Awaji, A. Takeuchi, M. Yabashi, H. Namatame, M. Taniguchi, H. Kumigashira, M. Oshima, M. Lippmaa, M. Kawasaki, H. Koinuma, K. Kobayashi, T. Ishikawa, and S. Shin, Phys. Rev. Lett. \textbf{93}, 236401 (2004).
		
\bibitem{chainani}
A. Chainani, M. Matsunami, M. Taguchi, R. Eguchi, Y. Takata, M. Oura, S. Shin, K. Sengupta, E. V. Sampathkumaran, Th. Doert, Y. Senba, H. Ohashi, K. Tamasaku, Y. Kohmura, M. Yabashi, and T. Ishikawa, Phys. Rev. B \textbf{89}, 235117 (2014).
		
\bibitem{CeCuSb2}
A. A. Abozeed, K. Sano, K. Terashima, A. Yamasaki, A. Higashiya, H. Fujiwara, T. Kiss, A. Sekiyama, Y. Tanaka, M. Yabashi, K Tamasaku, T. Ishikawa, S. Masubuchi, and S. Imada, JPS Conf. Proc., \textbf{30}, 011104 (2020).

\end{thebibliography}
\end{document}